\documentclass{article}

\usepackage{opex3}
\usepackage{graphicx}

\newcommand{\vect}{\textbf}

\begin{document}

\title{Nano-particle characterization by using Exposure Time Dependent 
  Spectrum and scattering in the near field methods:
  how to get fast dynamics with low-speed CCD camera.}

\author{Doriano Brogioli $^1$, Fabrizio Croccolo $^2$, Valeria Cassina $^1$, 
  Domenico Salerno $^1$, Francesco Mantegazza $^1$}

\address{$^1$
Dipartimento di Medicina Sperimentale,
Universit\`a degli Studi di Milano - Bicocca,
Via Cadore 48,
Monza (MI) 20052,
Italy.
}
\address{$^2$
Dipartimento di Fisica "G. Occhialini" and PLASMAPROMETEO,
Universit\`a degli Studi di Milano - Bicocca,
Piazza della Scienza 3,
Milano (MI) 20126,
Italy
}

\email{dbrogioli@gmail.com}

\begin{abstract}
     Light scattering detection in the near field, a rapidly expanding
     family of scattering techniques, has recently proved to be an
     appropriate procedure for performing dynamic measurements. Here
     we report an innovative algorithm, based on the evaluation of
     the Exposure Time Dependent Spectrum (ETDS), which makes it
     possible to measure the fast dynamics of a colloidal suspension
     with the aid of a simple near field scattering apparatus and
     a CCD camera. Our algorithm consists in acquiring static spectra
     in the near field at different exposure times, so that the measured
     decay times are limited only by the exposure time of the camera
     and not by its frame rate. The experimental set-up is based
     on a modified microscope, where the light scattered in the near
     field is collected by a commercial objective, but (unlike in
     standard microscopes) the light source is a He-Ne laser which
     increases the instrument sensitivity. The apparatus and the
     algorithm have been validated by considering model systems of
     standard spherical nano-particle. 
\end{abstract}


\bibliography{etds}

\section{Introduction}

     During the last few decades, light scattering techniques have
     demonstrated their ability to provide detailed information about
     properties of complex fluids. This capability is mainly due
     to the fact that these methods allow direct measurement of ensemble
     averages of the spatial and temporal fluctuations of fluid properties,
     which are precisely the quantities that can be usually calculated
     by theoretical methods of statistical mechanics, thus making
     available a simple way to compare experimental results with
     theory. 

     More recently, low angle light scattering methods have also became
     very popular thanks to the advent of pixilated sensors like
     CCD or CMOS cameras and their continuous technical development.
     The utilization of such sensors was first proposed by Wong and
     Wiltzius in 1993 \cite{wong1993} and gave rise to a large number of similar
     approaches \cite{ferri1997, cipelletti1999}. In these first attempts to utilize a multi-element
     sensor to gain statistical accuracy (typical sensors have about
     1k x 1k pixels, thus providing $10^6$ independent measurements at 
     one shot), the sensor was positioned
     in the Fourier plane of a lens. In this way it was possible
     to collect the scattered light in the far field, after removing
     the probe beam by means of a small mirror in the center of the
     focal plane. Accordingly, we will refer to this family of procedures
     as Scattering In the Far Field (SIFF) techniques. In order to
     overcome some of the difficulties experimented with SIFF, an
     alternative approach was recently proposed, consisting in collecting
     the light scattered by the sample in the near field, i.e. very
     close to the sample. We will refer to this second family of
     techniques as Scattering In the Near Field (SINF). The SINF
     methods show some advantages with respect to SIFF methods, including
     the extension of the scattering angle range to arbitrarily small
     values and the avoidance of stray light problems. Examples of
     SINF techniques are the shadowgraph \cite{wu1995, trainoff2002, brogioli2000},
     the schlieren, the
     speckle schlieren \cite{brogioli2003}, and the Heterodyne Near Field Scattering
     (HNFS) \cite{brogioli2002}. In the present paper, we will consider heterodyne
     SINF techniques only, in which both the scattered beam and the
     more intense transmitted beam are conveyed to the sensor. In
     this case, Fourier analysis of the heterodyne SINF images allows
     to recover the intensity of the scattered beams from the interference
     fringes they generate on the nearly uniform transmitted beam.
      
    SINF procedures have also been extended in the time domain in
     order to perform Dynamic Light Scattering (DLS) measurements
     \cite{croccolo2006, croccolo2006bis, croccolo2007, magatti2008},
     offering detailed data of the time correlation functions
     of the non-equilibrium fluctuations or of the diffusional random
     walk of colloidal particles. This approach has also been applied
     to optical microscopy \cite{cerbino2008}, suggesting that, even with incoherent
     illumination, one can achieve information about sample dynamics.
     
    When using SINF techniques the dynamic data are obtained by acquiring
     sequences of images at a given frame rate. After the acquisition,
     the images are Fourier transformed and processed. In the literature
     two different algorithms have been used to analyze the sequences,
     namely: (1) calculation of the time-correlation function (or
     equivalently the time power spectrum) and (2) calculation of
     the time structure function. In the present paper we introduce
     and describe a third method, which allows measuring fast dynamics
     without the need for a fast acquisition.

    Approach (1) is fairly similar to the standard procedure used
     in DLS experiments, where the time-correlation function is obtained
     by hardware calculation using the data collected by a photo-multiplier
     tube aligned at a certain angle with the probe beam \cite{berne1976}. For
     example, this approach has been used on shadowgraph images to
     investigate the dynamics of spiral defect chaos in high pressure
     CO$_{2}$ \cite{morris1996} and also to analyze the propagating modes that appear in
     a fluid subjected to a temperature gradient \cite{takacs2008}.

    Method (2) originated from the pioneering proposal by Schulz-DuBois
     \& Rehberg \cite{schulz-dubois1981} to develop hardware (a so
     called ``structurator'')
     capable of computing the structure function in place of the
     standard ``correlator''. This technique permits to eliminate the
     annoying slowly moving drifts that would plague scattering experiments.
     This was first reported by some of us 
     \cite{croccolo2006, croccolo2006bis}, and was later
     applied to a broad variety of different experimental methods
     like X-ray scattering \cite{cerbino2008bis}, 
     optical microscopy \cite{cerbino2008} and HNFS
     \cite{magatti2008}.

    The innovative approach proposed in the present paper is based
     on the study of the dependence of the (static) image power spectrum
     on the camera exposure time, hence the name Exposure Time Dependent
     Spectrum (ETDS). The physical idea behind this technique is
     that, when increasing the exposure time, the fast dynamics of
     the images is progressively faded out and what is left are the
     long time fluctuations only. In practice, increasing the exposure
     time is equivalent to the application of a sort of low pass
     filter, averaging the fastest fluctuations and thus depleting
     the power spectrum. For very long exposure times the fluctuations
     are completely smeared and we detect the background electronic
     noise of the CCD camera. Normally, state-of-the-art methods
     can detect fast dynamics up to the camera frame rate; by contrast,
     our ETDS method works by acquiring uncorrelated images at a
     very low frame rate, and the fastest dynamics it can detect
     is limited only by the shortest exposure time the camera can
     reach. This methodology intrinsically boosts the performance
     of any camera, since the shortest gating time is always much
     shorter than the fastest frame-rate delay. We suggest that the
     method can yield very impressive results, down to the sub-picosecond
     scale, by using pulsed lasers or fast-gated cameras or intensifiers.
     The method has been validated by measuring the time constants
     of three colloidal particles of different sizes using a HNFS
     set-up implemented on a microscope. 

    The rest of the paper is organized as follows: in section 2 the
     experimental apparatus used to collect scattering images will
     be described; in section 3 the ETDS will be derived; in section
     4 the experimental data for a wide range of wave vectors will
     be presented; finally, a brief summary of the work will be given.

\section{Experimental set up}
\begin{figure}
\begin{center}
\includegraphics{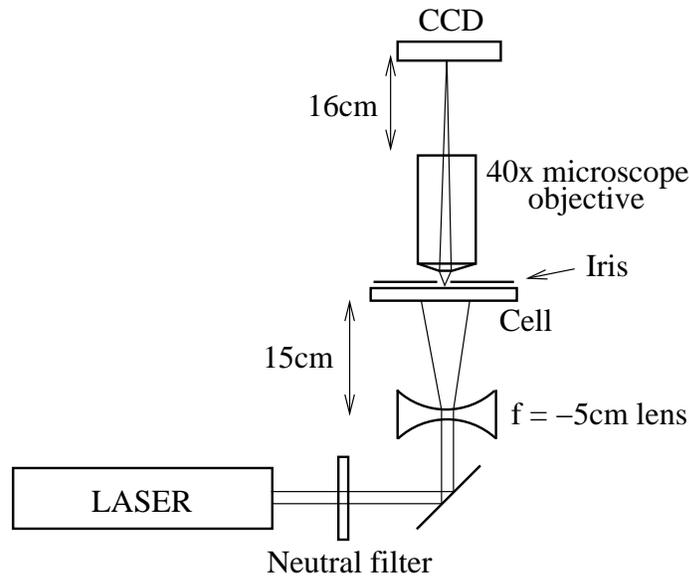}
\end{center}
\caption{
     Scheme of the optical set-up. A plane parallel beam of
     a He-Ne laser is attenuated by a neutral filter and bend upwards
     by a mirror. The beam is expanded by means of a negative focal
     length lens, making it diverge slightly before it enters the
     sample cell. Scattered light is acquired in the near field together
     with transmitted light, through a $40\times$ microscope objective,
     which conjugates a plane close to the sample onto the CCD sensor.}
\label{fig_setup}
\end{figure}

    In Fig. \ref{fig_setup} a sketch of the optical set-up is presented. It is based
     on a modified microscope where the light source is a He-Ne laser
     (10~mW, NEC) with an output of 0.7~mm diameter TEM00. The beam
     is used directly, without spatial filtering, because the quality
     of the unmodified beam has been proved to be enough for our
     purpose. The laser beam passes through a neutral filter wheel,
     with transmission range 0.3 - 0.0003. After the filter, the
     beam is reflected upward by a mirror, and goes through a -5~cm 
     focal-length, negative double-concave lens, in order to increase
     its diameter. We have checked that the slight divergence of
     the light impinging on the sample can be neglected. The sample
     is placed 15~cm after the lens, where the beam diameter is about
     2.5~mm wide. 

    The sample is placed in a glass cell with optical path 1 mm,
     made of two microscope cover slips, spaced by small glass strips
     cut from a microscope slide, and glued with silicone rubber.
     Over the top cover slip, we placed an iris with a 1 mm-diameter
     aperture, to reduce unwanted reflections. The optical system
     consists of a plan-achromatic 40x objective (Optika Microscopes,
     FLUOR), with 0.65 numerical aperture, 160~mm focal distance
     and 0.17~mm working distance; its focus lies on plane of the
     iris, about one mm outside the cell. Images are acquired with
     a CCD camera (Andor Luca), whose sensor is 658 x 496 pixels.
     The camera's maximum frame rate is 30 frames per second, with
     a minimum exposure time of 0.5 ms. The sensor is placed at 160
     mm from the microscope objective, so that it collects images
     directly with a magnification of about 40x. The CCD sensor images
     an area of $200 \times 150 \mathrm{\mu m^2}$, while the diameter 
     of the illuminated area of the sample is
     2.5~mm, thus providing the conditions for near field detection
     of the light \cite{magatti2008}. Each experiment consists of a set of measurements
     collected at different exposures times. For each measurement
     a different neutral filter is selected, and the exposure time
     is adjusted so that the acquired images have an average intensity
     corresponding to half of the dynamic range. One hundred images
     are then acquired, at a frame rate of 1~Hz, slowly enough to
     ensure that all the images are uncorrelated.

\section{Image processing}
    The goal of static scattering measurements is to obtain information
     about $I\left(\vect{Q}\right)$, the scattered intensity at transferred
     wave vector $\vect{Q}$, where 
     $\vect{Q}=\vect{k}_s-\vect{k}_i$, 
     $\vect{k}_s$ is the wave vector of the scattered beam
     and $\vect{k}_i$ is the wave vector of the incident 
     beam. In conventional static SIFF experiments, the static scattered intensity 
     $I\left(\vect{Q}\right)$ is obtained by direct measurement, 
     collecting the scattered
     light by using a photo-multiplier tube or a CCD in the far field
     at various angles with the impinging beam. By contrast, in a
     static SINF experiment using a pixilated sensor, we measure
     instantaneously the image signal 
     $i\left(\vect{x}\right)$, which is the intensity distribution 
     at point $\vect{x}$ on a plane close to the sample.
     After a set of images is acquired, the Image Power Spectrum (IPS) 
     $S_i\left(\vect{q}\right)$ is evaluated,
     calculating the average of the square modulus of the Fourier transform. 
\begin{equation}
S_i\left(\vect{q}\right)
= 
\left<\left|\tilde{i}\left(\vect{q}\right)\right|^2\right>
=
\left<\left|\mathcal{F}\left[i\left(\vect{x}\right)\right]\right|^2\right>
\end{equation}
    where $\mathcal{F}\left(\cdot\right)$ is the 2D Fourier transform, 
     $\vect{q}$ is the 2D spatial wave vector
     on the image plane, and the mean
     value $\left<\cdot\right>$ is obtained by averaging over the set of 
     images. Usually, before
     this Fourier processing, the images are ``cleaned'' by subtracting
     the optical background, which is obtained by averaging the images.
     The IPS $S_i(\vect{q})$
     represents the intensity of the Fourier modes, each one corresponding
     to a diffraction fringe generated by the interference between
     the transmitted beam and the scattered beam. By calculating
     the projection of $\vect{Q}$ on the image plane, it's easy to show 
     that the transferred wave vector $\vect{Q}$ is related to the 
     image wave vector $\vect{q}$ by the following relation:
\begin{equation}
Q\left(q\right) = \sqrt{2} k \sqrt{1-\sqrt{1-
\left(\frac{q}{k}\right)^2
}}
\end{equation}
where $k \approx k_s \approx k_i$ is the light wave vector in the medium. 
The relation between
the IPS $S_i\left(\vect{q}\right)$ and the scattered 
intensity $I\left[\vect{Q}\left(\vect{q}\right)\right]=I\left(\vect{q}\right)$
is linear, and is actually given \cite{croccolo2006, croccolo2006bis} by the 
sum of two terms:
\begin{equation}
S_i\left(\vect{q}\right) = 
I\left(\vect{q}\right) \cdot T\left(\vect{q}\right) +
B\left(\vect{q}\right)
\label{eq_transfer_function}
\end{equation}
    where $T\left(\vect{q}\right)$ is the instrument transfer function, i.e. a 
     relation between
     the wave vector and the instrument sensitivity, and $B\left(\vect{q}\right)$
     is the electronic background accounting for noise sources within
     the grabbing process. The instrument transfer function varies
     depending on the experimental set-up; it can be simply equal
     to a constant in the case of HNFS \cite{brogioli2002, ferri2004} or 
     schlieren \cite{brogioli2003, croccolo2006bis}
     thus providing the static scattered intensity with no further
     complications. Conversely, it has been shown 
     \cite{trainoff2002, brogioli2000, croccolo2007} that for
     a shadowgraph the transfer function exhibits, in some q range,
     deep oscillations, thus making the retrieval of scattered intensity
     difficult and providing no information in that vector range.
     In other cases the instrument transfer function can be completely
     unpredictable, which makes the static data completely unusable.
     However, the knowledge of the transfer function is not needed
     for the dynamic analysis, in which the attention is focused
     on the time fluctuations of the scattered intensity. 

    Indeed, the goal of dynamic light scattering is to study the
     time correlation function of the field $C_E\left(\vect{q},t\right)$.
     In the case of dynamic SIFF in heterodyne configuration, this
     quantity is usually directly measured with a correlator; or
     in the case of homodyne configuration, through Siegert's relation.
     Our ETDS-based algorithm for SINF analysis relies on the following
     argument. As pointed out by Oh et al. \cite{oh2004}, when the camera
     exposure time $\Delta t$ is not negligible, the IPS $S_i\left(\vect{q}\right)$
     depends also on $\Delta t$. Hence we will call it ETDS and use the notation
     $S_i\left(\vect{q},\Delta t\right)$. It is easy to find a relation between the 
     field correlation function $C_E\left(\vect{q},t\right)$ and 
     $S_i\left(\vect{q},\Delta t\right)$ as follows.
     An image $i\left(\vect{x},\Delta t\right)$, obtained with an exposure time 
     $\Delta t$, is the results of a time
     averaging over the instantaneous intensity map $i\left(\vect{x},t\right)$:
     $i\left(\vect{x},\Delta t\right) = \frac{1}{\Delta t} \int_{t_0}^{t_0+\Delta t} i\left(\vect{x},t'\right) \mathrm{d}t'$,
     where $t_0$ is the starting time of the image exposition. The same applies
     for the Fourier transform: $\tilde{i}\left(\vect{x},\Delta t\right) = \frac{1}{\Delta t} \int_{t_0}^{t_0+\Delta t} \tilde{i}\left(\vect{x},t'\right) \mathrm{d}t'$.
     The ETDS is then formally obtained as follows:
\begin{eqnarray}
S_i\left(\vect{q},\Delta t\right)
= 
\left<\left|\tilde{i}\left(\vect{q},\Delta t\right)\right|^2\right>
= \frac{1}{\Delta t^2}\left<
\int_{t_0}^{t_0+\Delta t}\mathrm{d}t'
\int_{t_0}^{t_0+\Delta t}\mathrm{d}t''
\tilde{i}\left(\vect{q},t'\right)
\tilde{i}\left(\vect{q},t''\right)
\right>
\nonumber
\\
= \frac{1}{\Delta t^2}
\int_{t_0}^{t_0+\Delta t}\mathrm{d}t'
\int_{t_0}^{t_0+\Delta t}\mathrm{d}t''
C_i\left(\vect{q},t'-t''\right)
\end{eqnarray}
    where $C_i\left(\vect{q},t\right)$ is the time correlation function of the 
    Fourier modes of the images. The double integral can be reduced to a simple
    integral with the following change of variables:
\begin{equation}
\begin{array}{l}
t' = t_0 + \frac{\Delta t}{2} + \frac{s'}{2} +\frac{s''}{2}\\
t'' = t_0 + \frac{\Delta t}{2} - \frac{s'}{2} +\frac{s''}{2}
\end{array}
\end{equation}
Substituting we have:
\begin{equation}
S_i\left(\vect{q},\Delta t\right) = 
\frac{1}{2 \Delta t^2}
\int_{-\Delta t}^{\Delta t}\mathrm{d}s'
\int_{-\Delta t + \left|s'\right|}^{\Delta t - \left|s'\right|}\mathrm{d}s''
C_i\left(\vect{q},s'\right)
\end{equation}
The calculation of the integral over $\mathrm{d}s''$ gives:
\begin{equation}
S_i\left(\vect{q},\Delta t\right) = 
\frac{2}{\Delta t^2}
\int_{0}^{\Delta t}\mathrm{d}s
\left(\Delta t-s\right)
C_i\left(\vect{q},s\right)
\end{equation}
but, in turn, $C_i\left(\vect{q},t\right)$ is related to $C_E\left(\vect{q},t\right)$
and so:
\begin{equation}
S_i\left(\vect{q},\Delta t\right) \propto
T\left(\vect{q}\right)
\frac{2}{\Delta t^2}
\int_{0}^{\Delta t}\mathrm{d}s
\left(\Delta t-s\right)
C_E\left(\vect{q},s\right)
+
B\left(\vect{q}\right)
\end{equation}
This is the general relation between $S_i\left(\vect{q},\Delta t\right)$ and the
correlation function $C_E\left(\vect{q},t\right)$: the ETDS calculated at a certain
exposure time $\Delta t$ is the average of the field correlation function, weighted by
a triangular function which vanishes for $s\to \Delta t$. This implies that spectra
obtained with different exposure times give different results, and the analysis of 
this variation as a function of $\Delta t$ brings information about the sample 
dynamics, or the field correlation function $C_E\left(\vect{q},t\right)$.
Now we consider the specific case we studied in our experiments:
colloidal particles performing a Brownian motion. In this case
the field correlation function is a decreasing exponential:
\begin{equation}
C_E\left(\vect{q},t\right) =
I\left(\vect{q}\right) \exp \left(-\frac{t}{\tau}\right)
\end{equation}
The resulting ETDS is:
\begin{equation}
S_i\left(\vect{q},\Delta t\right) \propto
I\left(\vect{q}\right) T\left(\vect{q}\right) 
f\left(\frac{\Delta t}{\tau}\right)
+
B\left(\vect{q}\right)
\label{eq_caso_esponenziale}
\end{equation}
where	
\begin{equation}
f\left(z\right) = 2 \frac{\exp\left(-z\right) -1 +z}{z^2}
\label{eq_f_exp}
\end{equation}
The last equation indicates that $S_i\left(\vect{q}, \Delta t\right)$
is a decaying function with a characteristic time $\tau$. In general this time
constant can be different for different wave vectors $\vect{Q}$. For example 
for diffusing Brownian particles the time constant varies as 
$\tau\left(Q\right) = 1/\left(DQ^2\right)$, in which $D$ is the translational 
diffusion coefficient of the particles in the solvent, as given by the 
Stokes-Einstein equation.
    
    It's worth analyzing the limits of the ETDS for very short and
     very long exposures times:
\begin{eqnarray}
\lim_{\Delta t /\tau \ll 1} S_i\left(\vect{q}, \Delta t\right) = 
I\left(\vect{q}\right) T\left(\vect{q}\right) + B\left(\vect{q}\right) 
\\
\lim_{\Delta t /\tau \gg 1} S_i\left(\vect{q}, \Delta t\right) = 
B\left(\vect{q}\right) 
\end{eqnarray}
   For very short exposure times the ETDS approximates the static
     value of the IPS as given by Eq. (\ref{eq_transfer_function}).
     For long times the ETDS goes asymptotically to the electronic background of the system.

    In summary, by acquiring series of images with many different
     exposure times and analyzing the ETDS for each single wave vector
     by using the fitting procedure in Eq. (\ref{eq_caso_esponenziale}) 
     and (\ref{eq_f_exp}), one obtains
     a direct and simultaneous measurement of the time constants
     of the sample for all the measured wave vectors. As a byproduct,
     one gets also two other terms from the fitting procedure, namely,
     the product of the static scattered intensity times the instrument
     transfer function $I\left(\vect{q}\right)\cdot T\left(\vect{q}\right)$ 
     and the electronic background of the system.

\section{Experimental results}
\begin{figure}
\begin{center}
\begin{tabular}{cc}
\includegraphics{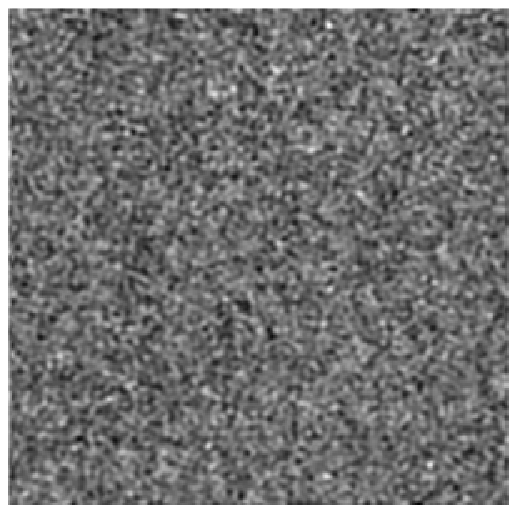} & \includegraphics{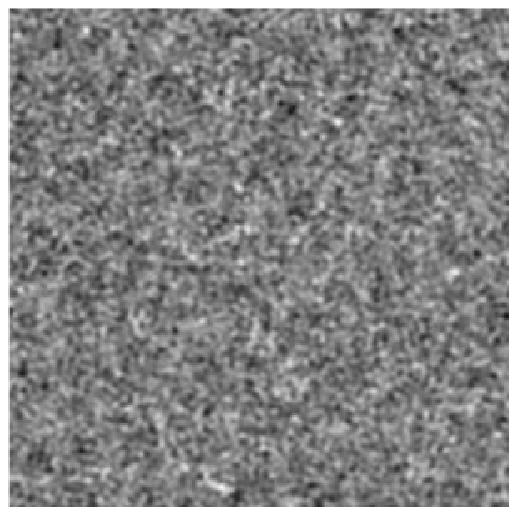}\\
a) & b)\\
\includegraphics{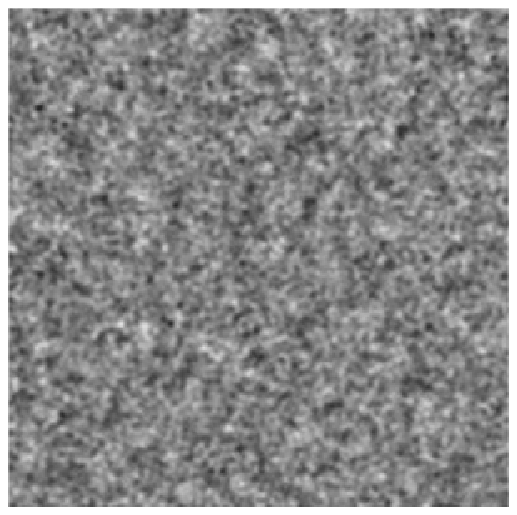} & \includegraphics{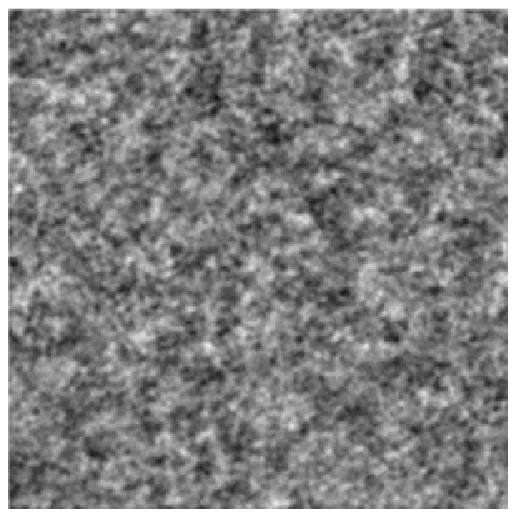}\\
c) & d)\\
\end{tabular}
\end{center}
\caption{Acquired images of nano-particles of 80~nm diameter at
     four different exposure times $\Delta t$ (a) 18~ms, b)~55 ms, c) 180~ms,
 d) 550~ms). The image size corresponds to $75\times 75 \mu m^2$ in real space.}
\label{fig_speckles}
\end{figure}

    In order to test the validity of the our approach we have performed
     three experiments with different commercial samples of polystyrene
     nano-particles of 80~nm, 150~nm and 400~nm diameter (Polyscience Inc.).
     The particles are well mono-dispersed and the declared size
     corresponds to the measured size within 5\% (81$\pm$4~nm, 149$\pm$5~nm, 402$\pm$12~nm
     respectively) as checked by a traditional DLS apparatus (Brookhaven
     Instruments, ZetaPlus). The two smallest samples have been dispersed
     in water at a concentration of 0.1\% w/w, whereas the largest
     ones have been prepared in water at a concentration of about
     0.001\% w/w. All the samples gave a sufficiently high scattering
     signal and for each sample, various sequences of 100 images
     were acquired at 7 different exposure times $\Delta t$ of about 
     0.5~ms, 1.5~ms, 5.5~ms, 18~ms, 55~ms, 180~ms, and 550~ms. For each sequence,
     the ETDS was calculated. The optical background due to stray
     light contributions was removed from each image by subtracting
     the image average obtained over the set of images for each exposure
     time. In Fig. \ref{fig_speckles} we show the images obtained with the 80 nm particles
     at four different exposure times. As it is apparent from the
     figures, as the exposure time increases the appearance of the
     images changes, and the short scale inhomogeneities are progressively
     smeared.

\begin{figure}
\begin{center}
\includegraphics[scale=0.4, angle=-90]{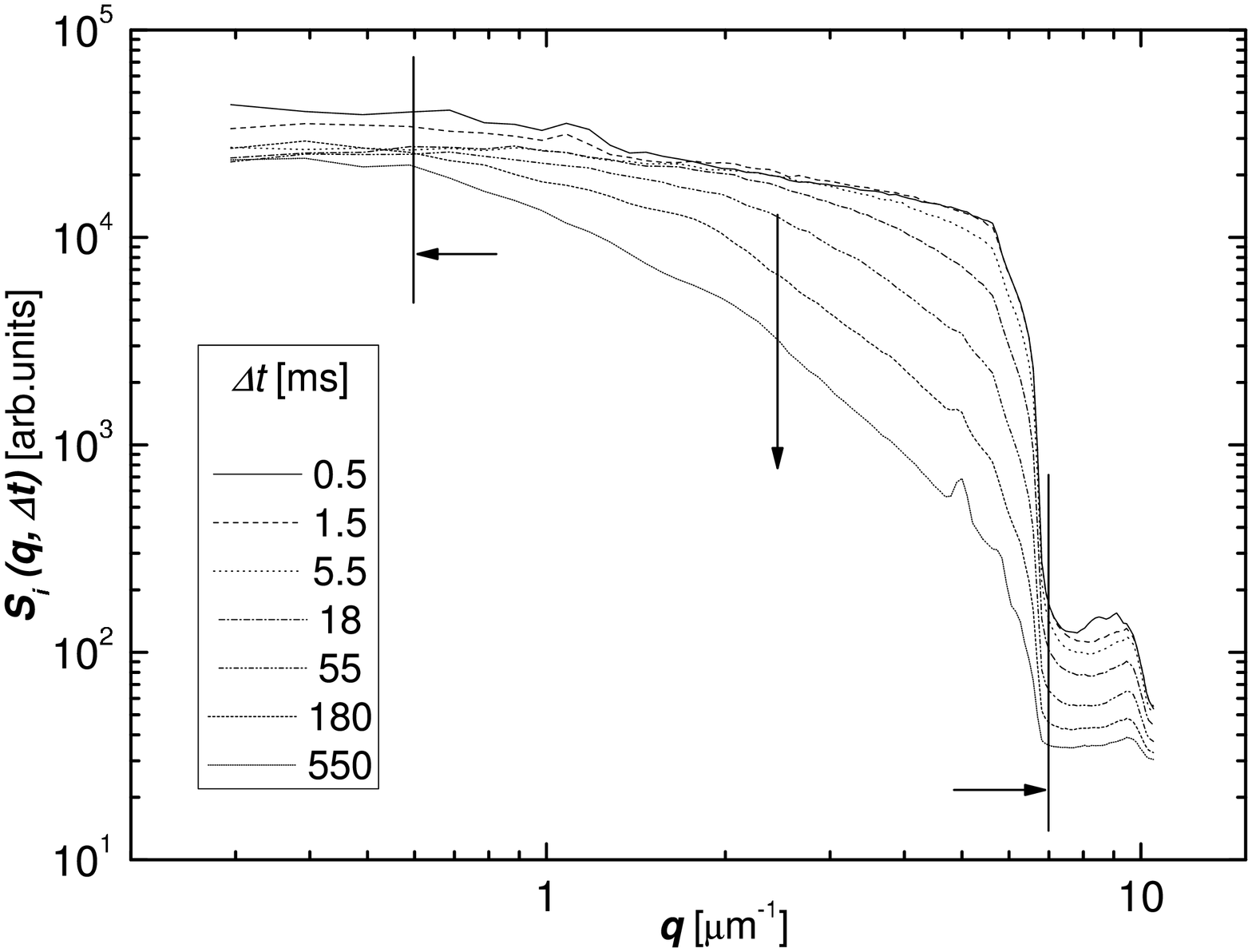}
\end{center}
\caption{Exposure Time Dependent Spectra $S_i\left(\vect{q},\Delta t\right)$
 measured as a function of the scattering vector $\vect{q}$, at different exposure 
 times $\Delta t$ for particles of 80~nm diameter. The central arrow indicates
     the increase of the exposure time, while the vertical lines
     indicate the lower and upper boundaries of the $\vect{q}$ range over which 
  the fitting procedure has been performed.}
\label{fig_spettri}
\end{figure}

    In Fig. \ref{fig_spettri} the $S_i\left(\vect{q},\Delta t\right)$ spectrum
     for 80~nm particles is shown at different exposure
     times. This data show clearly that the spectra decrease as the
     exposure time is increased. For the longest exposure time the
     averaging of the fluctuations reduces the scattering signal
     towards the electronic background level.

\begin{figure}
\begin{center}
\includegraphics[scale=0.4, angle=-90]{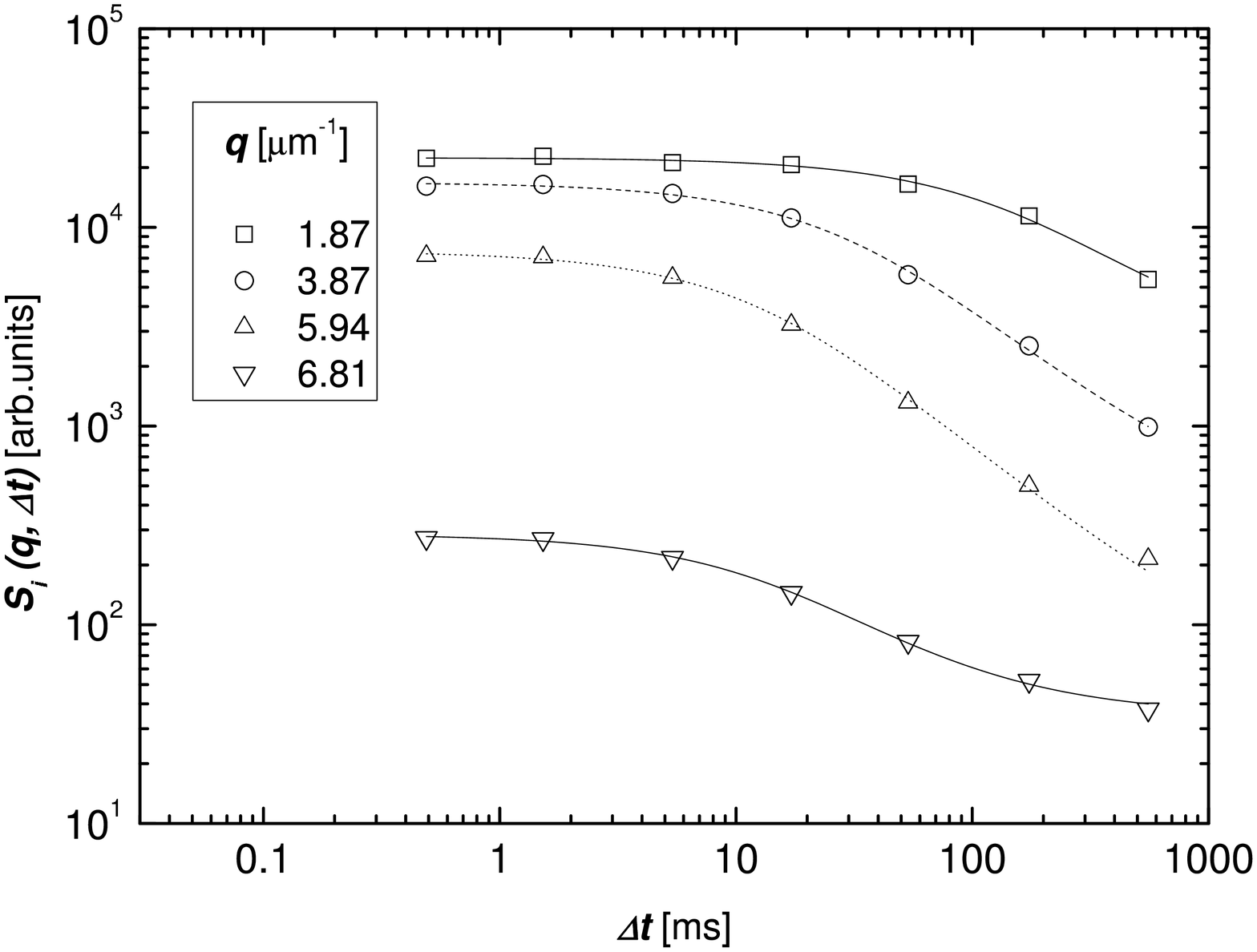}
\end{center}
\caption{Exposure time dependent spectra $S_i\left(\vect{q},\Delta t\right)$ 
   measured as a function of the exposure times $\Delta t$, at four different wave
   vectors $q$, for particles of 80~nm diameter.
     Lines correspond to fitting curves determined by Eq. \ref{eq_caso_esponenziale}
     and \ref{eq_f_exp} see the main text for details.}
\label{fig_varie_etds}
\end{figure}

    In Fig. \ref{fig_varie_etds} the $S_i\left(\vect{q},\Delta t\right)$ spectra are 
     represented as a function of the exposure time at 4 different wave vectors.
     It's possible to observe the different
     decay times and the different amplitudes of the electronic background.
     Fitting curves obtained by using equations 4 and 5 are displayed,
     too. The fitting is carried out with three free parameters,
     namely, the time constant $\tau\left(\vect{q}\right)$, the product of the 
     static scattered intensity times the instrument
     transfer function $I\left(\vect{q}\right) \cdot T\left(\vect{q}\right)$,
     and the electronic background of the system $B\left(\vect{q}\right)$. The agreement
     between data points and the corresponding fitting
     curves is very good. It's worth pointing out that, even if this
     fitting is made with such a limited number of data-points, each
     point is the output of a statistical analysis of 100 images
     and several statistically independent samples. is the number
     of independent samples is about $n \approx \pi q/q_{min}$, where $q_{min}$
     is the minimum wave vector. This means that in the wave vector
     range of interest the number of statistical samples for each
     point in Fig. \ref{fig_varie_etds} varies from 3,000 to 30,000.

\begin{figure}
\begin{center}
\includegraphics[scale=0.4, angle=-90]{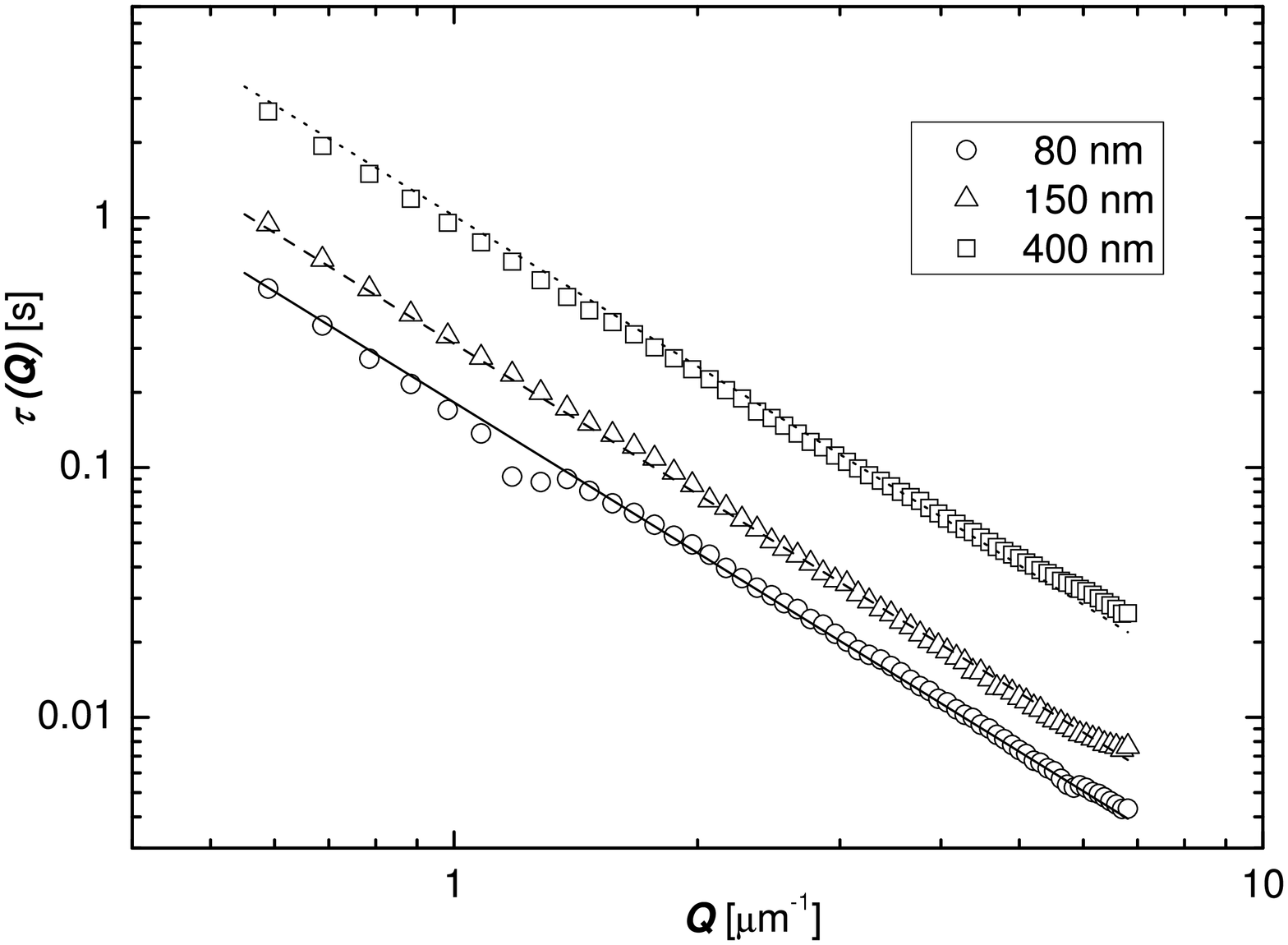}
\end{center}
\caption{Time constant $\tau$ measured as a function of the wave vector
     $q$, for 3 different particle sizes. Straight lines in the log-log
     plot correspond to fitting curves with the power law 
     $\tau\left(Q\right)=1/\left(DQ^2\right)$. }
\label{fig_tempi}
\end{figure}

    In Fig. \ref{fig_tempi} the time constants obtained by the proposed fitting
     procedure are plotted for the different samples as a function
     of the wave vector. Lines correspond to fitting of $\tau$ values
     according to the formula $\tau\left(Q\right)=1/\left(DQ^2\right)$. 
     This fitting procedure 
     has only $D$ as an adjustable parameter; the particle diameter can be recovered
     from the Stokes-Einstein equation. Obtained diameters are 81$\pm$2~nm,
     138$\pm$7~nm and 427$\pm$12~nm, in very good agreement with the sizes measured
     by standard DLS. 

\section{Conclusions}
    In the present work we have presented results from a dynamic
     near field scattering experiment, performed on colloidal particles
     by means of a simple set-up, consisting of a suitable modified
     optical microscope and a laser beam as illuminating light. A
     new statistical algorithm is presented that allows us to measure
     the characteristic time constants of the colloidal system. The
     statistical procedure enables the camera to detect characteristic
     times much shorter than the camera delay times. Indeed, the
     only limitation is associated with the minimum exposure time
     of the camera and not by its delay time. This result opens the
     way to the investigation of ultra-fast dynamics such as molecular
     motions, capillary waves, nano-particles diffusion, and virus
     or biological macromolecules mobility. Until now these phenomena
     required sophisticated equipment, whereas the proposed method
     uses a simple modified microscope and a standard camera.

\section*{Acknowledgments}

   This work has been supported by the financial funding of the
     EU (project BONSAI LSHB-CT-2006-037639 and project NAD CP-IP
     212043-2).

\end{document}